# Traffic Simulation in Ad Hoc Network of Flying UAVs with Generative AI Adaptation

Andrii Grekhov[1], grekhovam@gmail.com, ORCID: 0000-0001-7685-8706
Volodymyr Kharchenko[2], kharch@nau.edu.ua, ORCID: 0000-0001-7575-4366
Vasyl Kondratiuk[3], kon_vm@ukr.net, ORCID: 0000-0002-5690-8873

[1,2,3] Research Training Center "Aerospace Center", State University "Kyiv Aviation Institute", Kyiv, Ukraine

* Correspondence: grekhovam@gmail.com; Tel.: +38-098-286-5730

**Abstract:** The purpose of this paper is to model traffic in Ad Hoc network of Unmanned Aerial Vehicles (UAVs) and demonstrate a way for adapting UAV communication channel using Artificial Intelligence (AI). The modeling was based on the original model of Ad Hoc network including 20 UAVs. The dependences of packet loss on the packet size for different transmission powers, on the packet size for different frequencies, on the UAV flight area and on the number of UAVs were obtained and analyzed. The implementation of adaptive data transmission is presented in the program code. The dependences of packet loss, power and transaction size on time during AI adaptation are shown.

## Introduction

The development of the UAV field in recent years has shown that UAVs will be an integral part of future networking and communication systems [1-7]. Research is looking at UAV solutions in various fields, but a comprehensive study of AI-based autonomous UAVs has not yet been fully established [8-14].

The Ad Hoc topology (Fig. 1) is defined by a decentralized network scheme in which each UAV is able to communicate directly with any other node. There is no central node, and nodes can pass messages to other nodes, creating a self-healing network. The Ad Hoc topology is highly fault-tolerant, scalable, and flexible. Therefore, it is well suited for complex UAV networks where reliability and redundancy are critical [1, 2, 5].

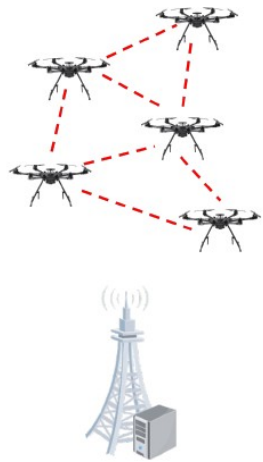

**Fig. 1**. UAV Ad Hoc network [10]

AI can potentially improve each of the key performance metrics of UAV communication links. AI algorithms can optimize UAV coordinates and movement for maximum coverage, taking into account obstacles, signal strength, and power consumption. AI can improve throughput by dynamically allocating resources, optimizing communication protocols based on data traffic patterns. Latency is an important metric for real-time applications, and AI can reduce latency by predicting and proactively reducing potential delays and optimizing data routes. AI can improve reliability by using error corrections, predicting and reducing potential failures [6-9].

Advances in AI are believed to be driven by machine learning techniques that enable systems to acquire knowledge from data and learn from experience [10]. AI has already moved from hand-coding to learning from big data.

The use of AI in UAVs has opened up new possibilities in surveillance, search and rescue, and military operations. Intelligent drones can scan areas, perform thermal imaging, or reconnaissance. AI has made it possible for drones to operate autonomously and perform complex tasks. AI-powered UAVs can make informed decisions while flying, and this autonomous capability is especially useful in military drones, where quick, real-time responses are critical. AI has made swarm technology possible, where multiple drones can work in a coordinated manner. In the military defense arena, AI-powered drones have become key. The use of AI in UAVs has changed military operations, allowing them to be used more complexly and strategically in reconnaissance, surveillance, and combat scenarios. The war in Ukraine is an example of how AI-powered UAVs are changing military strategies. AI and machine learning have made drones an integral part of advanced military tactics. AI-powered UAVs in military operations leverage the confluence of technological innovation and strategic application, bringing profound impact of AI to the defense sector [15, 16].

The operation of UAV Ad Hoc network is impossible without reliable communication channels between all components. This study is developing methods for assessing the parameters of real traffic in UAV Ad Hoc network and adapting UAV communication channel parameters using AI. Our contributions can be summarized as follows. The dependences of packet loss on the packet size for different transmission powers, on the packet size for different frequencies, on the UAV flight area and on the number of UAVs were obtained and analyzed. The implementation of adaptive data transmission is presented in the program code. The dependences of packet loss, power and transaction size on time during AI adaptation are shown.

Our contributions can be summarized as follows. The dependences of packet loss on the packet size for different transmission powers, on the packet size for different frequencies, on the UAV flight area and on the number of UAVs were obtained and analyzed. The implementation of adaptive data transmission is presented in the program code. The dependences of packet loss, power and transaction size on time during AI adaptation are shown. The results obtained are of practical importance, since they allow predicting the behavior of UAV Ad Hoc network and adapting it.

This article is organized as follows. In *Related Works*, we review works related to UAV Ad Hoc network. In *Model and Method of Calculation* we describe the architecture of proposed model and details of calculations. In *Results of Simulation,* we describe obtained data. In *Implementation of Adaptive Data Transfer,* we demonstrate the way of adapting UAV communication channel parameters using AI. In *Conclusion*, we note the contribution of our research to the development of methods for predicting the behavior of UAV Ad Hoc network and adapting it.

**Related Works**
In [1], the use of UAV as an intelligent repeater to improve communication in a wireless mesh network is investigated. The UAV moves autonomously to maximize communication based on a positioning algorithm that uses radio measurements collected in the network. The performance of the developed algorithm is tested in a real experiment.

For UAV interaction, it is necessary to create a network communication architecture with the implementation of suitable routing protocols. The article [2] considers network communication architectures and various ad-hoc routing protocols that fall under proactive, reactive and hybrid protocols. A network communication architecture based on the clustering concept and an original idea of the routing mechanism is proposed.

Using the right networking strategies for UAV swarms allows users to communicate over distances much greater than in terrestrial applications. In [3], a hybrid communication architecture for UAV swarms using a heterogeneous radio network is proposed. It is based on long-range communication protocols such as LoRa and LoRaWAN and IEEE 802.11s protocols. The advantages of such an architecture, limitations, possible implementation and use cases are discussed.

The paper [4] provides an overview of UAV networks, communication channels, routing protocols, mobility models, research problems and simulation software for FANET. The routing protocol for topology-based FANET is discussed in detail with detailed categorization, descriptions and qualitative comparative analysis. The research topics and problems that need to be solved before UAV communication is expected to become a reality and practical in the industry are described.

The article [5] provides an overview of the different types of UAVs used in FANET, their mobility models, main characteristics and applications. Routing protocols and new technologies integrated with FANET are considered.

The review [6] categorized the works based on three different schemes: 1) based on the application scenario, 2) based on the AI algorithm, and 3) based on the AI learning paradigm. A selection of frameworks, tools, and libraries used in UAV systems integrated with AI is presented. It is noted that the integration of AI in UAVs includes issues ranging from path planning to resource allocation. Reinforcement learning-based algorithms are more often used in UAV systems than other AI algorithms.

The objective of the paper [7] is to review AI-based autonomous UAV networks. More than 100 papers on UAVs are reviewed with a focus on the classification of autonomous functions, network resource management and scheduling, multiple access and routing protocols, and power management and energy efficiency for UAV networks. It is concluded that AI-based UAVs are a technologically feasible and economically viable paradigm.

Chapter [8] discusses UAV networks with the integration of aerial robotics, AI, and wireless communications. The conceptualization, development, applications, and challenges associated with UAVs are analyzed. The need to address regulatory, privacy, ethical, and technical issues is emphasized. It is noted that the development of UAVs depends on continuous innovation in drone technology, AI algorithms, and communication infrastructures. The confluence of AI and UAVs will have a profound impact on industries and society, shaping the future of automation and communications.

The article [9] is devoted to intelligent Software-Defined Networks (SDN) and UAVs. An analysis of the evolution from traditional network environments to UAV-based SDN environments is presented. The performance, security, latency and communication efficiency of UAVs are investigated. A taxonomy, comparison and analysis of existing machine learning solutions specifically designed for UAV-based software-defined networks are considered.

The review [10] examines AI applications in UAV-enabled wireless networks. The background and motivation for AI integration are described, highlighting the potential for improving network performance, autonomy, and adaptability. AI applications include data collection and processing, trajectory placement and optimization, radio resource management, routing and topology management, edge computing and caching, and enhancing security and privacy. The integration of AI with UAV-enabled wireless networks (UWNs) has great potential to transform wireless communications.

Study [11] analyzes AI-based autonomous UAV networks. Classification of autonomous capabilities, provisioning of network resources, network planning and selection, multiple access and routing protocols, power management, and energy consumption strategies in UAV networks are considered. It is concluded that AI-based UAVs are a profitable and technologically feasible choice for future networks.

The main issue in developing UAV swarm applications is the communication system shared between flying drones and ground base stations. Proper networking strategies for UAV swarms allow users to exchange data over distances much greater than in ground-based applications. In [12], a hybrid communication architecture for UAV swarms is proposed using a heterogeneous network based on long-range communication protocols such as LoRa and LoRaWAN and IEEE 802.11s protocols.

In [13], a modeling methodology based on training data of a generative neural network is presented. The proposed generative model first predicts the link state (line of sight, no line of sight, or disconnected) and then feeds this state into a variational autoencoder that generates path loss, delay, and angle of arrival. The methodology is demonstrated for 28 GHz air-to-ground links between a UAV and a cellular system in urban environments with training datasets. Both street-level ground base stations and rooftop air base stations were used.

Generative Artificial Intelligence (GAI) is used to solve problems in wireless communication scenarios. The paper [14] discusses the applications of GAI to improve the communication and network performance of UAVs. GAI technologies and the roles of UAV networks are reviewed. It is shown how GAI can improve the communication, network, and security performance of UAV systems. A new GAI framework for UAV networks is proposed and an example of data rate optimization based on it is presented.

The structural diagram of UAV adaptive channel with AI is proposed in [15]. The method of adapting the communication channel is demonstrated based on the model "Ground Control Station – Stratospheric Repeater – UAV", which was built using the MATLAB Simulink software. The dependences of the Bit Error Rate (BER) on the Signal-to-Noise Ratio (SNR) for various data transmission rates used for GAI are obtained and analyzed.

The article [16] shows a method for adapting UAV communication channel using AI. The simulation was carried out using the model "Ground Control Station - Satellite - Air Repeater - UAV", which was built using the NetCracker software. Adaptation of parameters depending on the delay and the number of bit errors was carried out using a linear regression model. Adaptive transmission was implemented based on predicting the packet size for a given delay and dynamically changing the packet size.

**Model and Method of Calculation**

To simulate traffic in an ad hoc network of flying drones, we created a model (Fig. 2) of 20 UAVs for which random coordinates are generated in a given area (1500x1500 meters) and 10 source/receiver pairs are created for data transmission (Appendix 1).

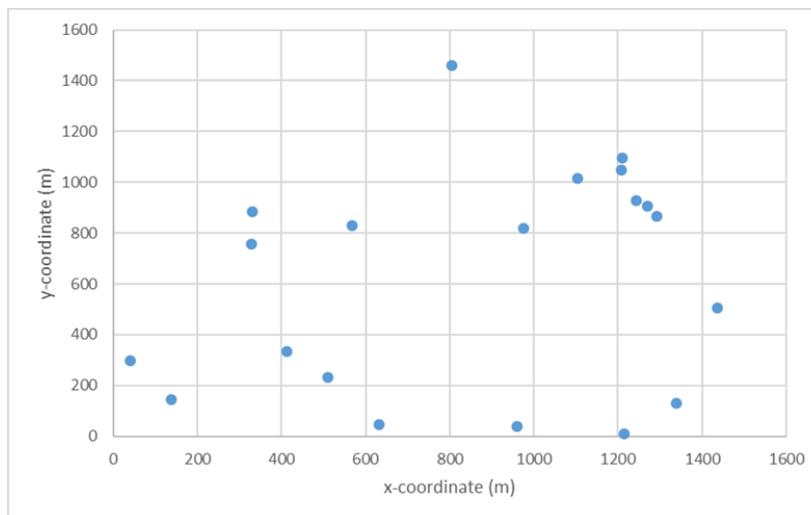

**Fig. 2** Ad-Hoc network of 20 UAVs

*Loss calculation*

The Friis loss model [17] is used to calculate the packet loss rate depending on the distances between devices and transaction parameters. The Friis equation for free-space path loss is:

$$\text{FSPL} = 20\log_{10}(d) + 20\log_{10}(f) + 20\log_{10}(4\pi/c),$$

where *d* is the distance between the transmitter and receiver in meters, *f* is the signal frequency in Hertz and *c* is the speed of light ($\approx 3\cdot 10^8$ m/s).

The Friis transmission equation describes the power received by an antenna under ideal Line-of-Sight (LOS) conditions, primarily in free space. It is widely used in communication systems, including UAV-based links, to calculate the signal strength at a given distance.

The loss increases as $20\log 10(f)$, which is 6 dB per doubling of frequency. This is because the wavelength decreases with increasing frequency, making the signal less resistant to scattering and absorption. The loss increases as $20\log 10(d)$, which is 6 dB per doubling of distance. This reflects the decrease in power delivered to the receiver due to the energy being dispersed over a larger area. Higher frequencies and longer distances result in greater signal attenuation.

The model assumes an ideal free space environment, where no obstructions, reflection, diffraction, or scattering affect the signal. For UAV links, this assumption works well in high-altitude scenarios, but deviations may occur close to the ground or in urban areas. The Friis model is only valid for LOS conditions. In the presence of obstacles, the model underestimates the actual path loss. UAVs often maintain LOS communications, making this model relevant for estimating the theoretical link budget.

The model does not account for real-world phenomena such as multipath fading, rain fading, or vegetation loss, which are critical for UAV communications under certain conditions. The model does not account for interference from other sources that can affect communications, especially in crowded spectral bands. Nevertheless, the Friis loss model provides a baseline for estimating path loss in UAV networks, helping to design link budgets and determine the effective range.

*Estimating the probability of packet loss*

The formula *loss_prob = 1 - (1 - BER) \*\* packet_size* is used to estimate the probability of losing a data packet during transmission over a communication channel. *BER* is the probability of a single bit being corrupted during transmission. A packet consists of multiple bits. The total number of bits in a packet is determined by packet_size. Assuming independent bit errors, the probability that a single bit will be uncorrupted is *1−BER*. The probability that all packet_size bits in a packet will be error-free (no bits are corrupted) is: *(1 - BER) \*\* packet_size*. This assumes that bit errors are independent and identically distributed, which is often a reasonable assumption in channels dominated by random noise. The probability of packet loss is the probability that at least one bit in a packet will be corrupted. It is determined by the formula: *loss_prob = 1 - (1 - BER) \*\* packet_size*.

This formula is commonly used in communication systems to estimate the reliability of data transmission for given BER values and packet sizes. It highlights the exponential relationship between packet size and loss probability: as packet size increases, even small BER values can lead to significant packet loss probability.

If the *SNR* is expressed on a linear scale, then the following [18] approximation can be used for an approximate calculation:

$$BER \approx \frac{1}{2}e^{-SNR},$$

where *SNR* is the signal-to-noise ratio.

**Results of Simulation**

The results calculated for the models and presented in this section are very important for understanding the behavior and operation of the Ad-Hoc network of UAVs. Their value lies in the fact that they are almost impossible to obtain experimentally.

The disadvantages of any models are associated with the impossibility of reproducing the features of the modeled system with high accuracy. Therefore, if there are no experimental data with which the calculated values could be compared and, thus, confirm the adequacy of the model, then such modeling is not trusted too much and requires justification of the applicability limits. In this case, one can expect that the changes in values calculated within the same model will be adequate, as if the shortcomings of the model itself were eliminated.

The Python code creates an Ad-Hoc network of 20 UAVs within a given area and calculates the packet loss for different transaction sizes using the Friis loss model (Appendix 1). The loss is calculated taking into account the distances between UAVs and transmission characteristics: packet size, transmission power, successful transmission threshold, frequency, number of UAVs, and the size of the given area.

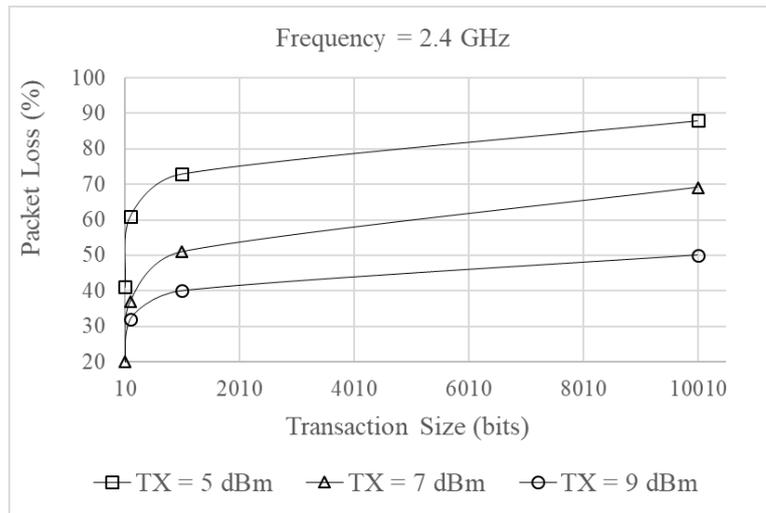

**Fig. 3** Dependences of Packet Loss on TS for different transmit power

Fig. 3 shows the dependences of packet losses on the transmission power. It is clear that with decreasing power, the losses increase. But one interesting regularity is noteworthy, which obviously can be generalized to other power ratios. In the range of (100, 1000, 10000) bits, the losses increase by (1.89, 1.83, 1.76) times, respectively, when moving from 9 dBm to 5 dBm (9:5=1.8). Similarly, when moving from 7 dBm to 5 dBm (7:5=1.4), the losses increase by (1.6, 1.4, 1.3) times, respectively, and when moving from 9 dBm to 7 dBm (9:7=1.3), the losses increase by (1.2, 1.3, 1.38) times, respectively. Since the average increase in losses in the cases considered is 1.82, 1.43 and 1.29 times, respectively, an empirical rule can be formulated: on average, losses increase as many times as the power decreases.

Losses increase logarithmically with frequency (Figure 4). Therefore, in practical applications, frequencies such as 28 GHz (used in 5G) will have much higher FSPLs than lower frequencies such as 2.4 GHz (used in Wi-Fi or Bluetooth). Higher frequency signals are more susceptible to packet loss. Lower frequency signals are better suited for long-range communications due to their lower FSPL, but may suffer from interference due to crowding in commonly used bands.

Calculations have shown that correlations similar to the previous case are absent here. As the frequency increases, the losses also increase. Figure 4 shows the results for a fixed power. It is obvious that increasing the power will lead to a decrease in losses in the entire range of frequencies and packet sizes.

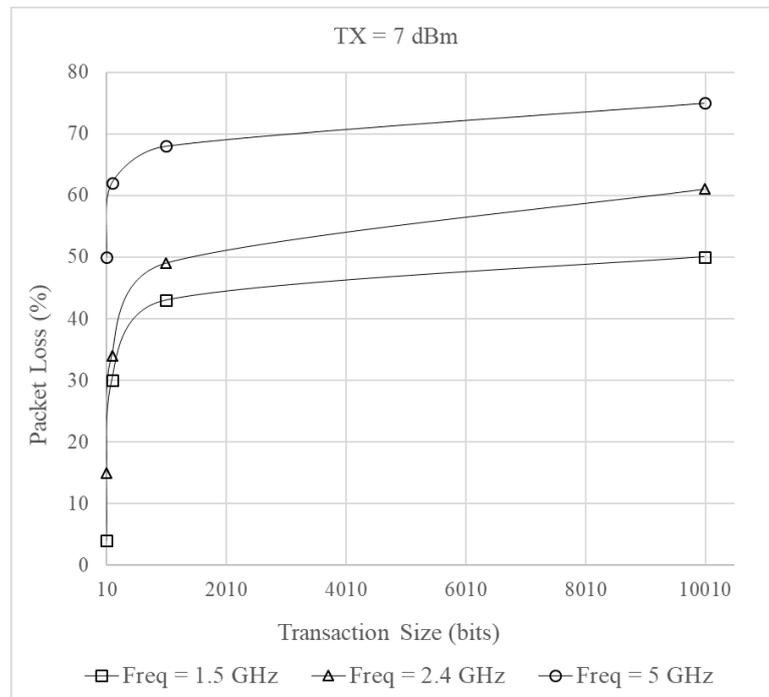

**Fig. 4** Dependences of Packet Loss on packet size for different frequencies

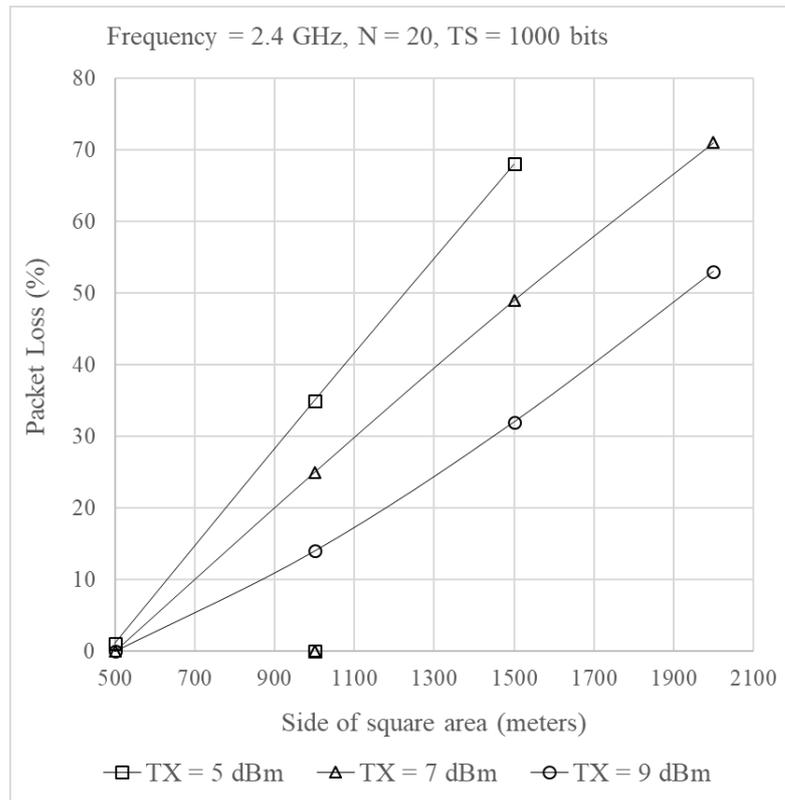

**Fig. 5** Dependences of Packet Loss on UAV flying area

Packet loss (Fig. 5) in Ad-Hoc UAV networks is highly dependent on the size of the area in which they operate. As the area size increases, UAVs become more dispersed, increasing the average distance between nodes. This increased distance can result in greater signal attenuation and a decrease in signal-to-noise ratio, especially in environments with free-space path loss models such as Friis. This is important because the probability of successful packet delivery decreases with weaker signals, leading to higher packet loss rates.

In large areas, topology changes occur and UAVs are more likely to move out of range of their neighbors. This leads to frequent disconnections and reconfigurations of routing protocols. This is important to consider because routing protocols need time to find new routes when connections are broken. During this time, packets may be lost.

Large areas typically result in multi-hop communications, since direct connections between sources and destinations are less likely. Each hop introduces the potential for interference, queuing delays, and packet loss due to increased collision or buffer overflow probability.

For a fixed number of UAVs, larger areas reduce network density, and sparse networks have fewer available neighbors to route to, which can lead to increased reliance on distant nodes or isolated nodes unable to communicate.

Larger areas can lead to overlapping transmissions in regions with higher UAV concentrations or during convergence zones where nodes are temporarily grouped together. This causes packet collisions and increased retransmissions, especially in protocols that rely on contention-based access, such as Wi-Fi.

At the same time, smaller areas have higher network density and shorter distances between nodes, which leads to reduced signal loss due to topology changes.

The practical implications of our results (Fig. 5) are that network performance in UAV swarms must balance trade-offs between area size, communication range, and number of UAVs. Adaptive routing protocols, efficient transmit power control, and UAV movement coordination can help mitigate the effects of packet loss over large areas.

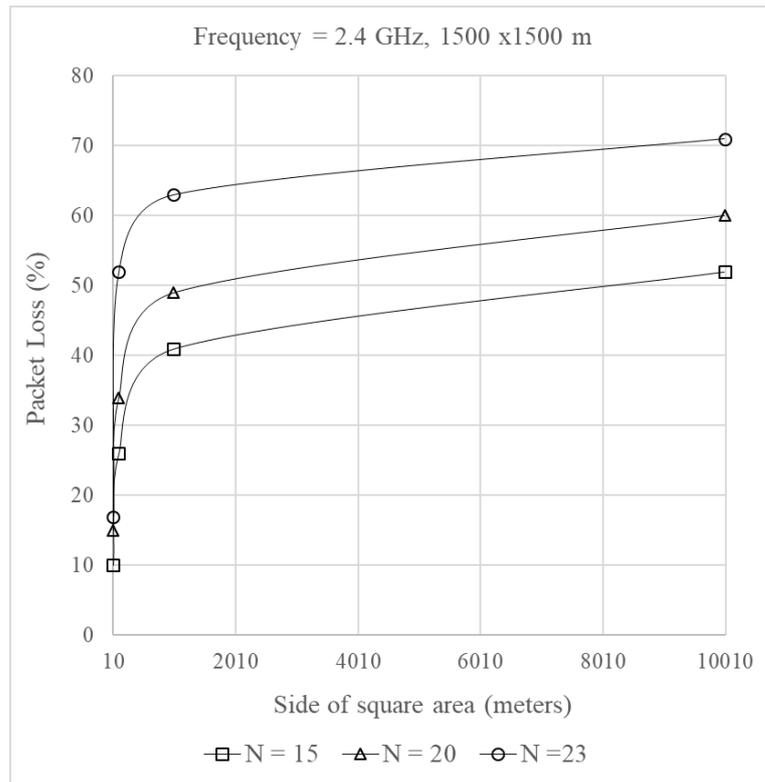

**Fig. 6** Dependences of Packet Loss on UAV number

Packet loss (Fig. 6) in Ad-Hoc UAV networks also depends on the number of UAVs in a fixed area. As the number of UAVs increases, the network density increases. This results in more potential communication channels, but also more interference and contention for the wireless channel. In high-density networks, packet collisions occur more frequently due to simultaneous transmissions, especially in contention-based access protocols such as Wi-Fi (802.11). This increases packet loss.

With more UAVs transmitting data, the wireless channel becomes congested, which increases contention for bandwidth. Congestion results in higher latency, buffer overflow, and packet loss, especially if the traffic exceeds the channel capacity.

Denser networks require more routing updates and control message exchanges. Protocols generate more frequent route discovery and maintenance messages as the number of UAVs increases. Control messages consume bandwidth and increase the likelihood of collisions, indirectly contributing to packet loss.

In dense networks, routes often have more hops due to the large number of intermediate nodes. Additionally, the mobility of UAVs can result in frequent connection breaks and path reopenings. Each additional hop increases the chance of packet loss due to interference, errors, or buffer overflows.

As the number of UAVs increases, the same frequency band can be reused more often, causing interference between adjacent connections. This reduces the effective signal-to-noise ratio, resulting in higher packet error rates.

This can be summarized as follows. At low UAV counts, the network is underutilized, with fewer opportunities for collisions or congestion, resulting in low packet loss. However, sparse networks can suffer from connectivity issues. The optimal number of UAVs can balance connectivity and manageable interference.

At high UAV counts, dense networks face severe congestion, interference, and routing overhead, which significantly increases packet loss.

The practical implications of our results (Fig. 6) are that dynamic resource allocation (e.g., adaptive transmit power, channel selection) and congestion control mechanisms should be implemented to optimize performance. Proper tuning of the number of UAVs depending on the area size and communication protocol is critical to minimize packet loss.

**Implementation of Adaptive Data Transfer**

Let us divide the task into several stages. First, we will write code that will predict the packet size based on the given losses and signal power (Appendix 2), and then we will model the adaptation of the signal power from the transmission parameters to reduce the level of losses (Appendix 3).

Today, AI methods rely on known data for training and decision making. However, collecting high-quality data can be challenging due to the dynamic and complex nature of these networks. In the field of communications and UAV networks, obtaining large amounts of well-labeled data is difficult. Ensuring sufficient data availability and quality is critical to training effective AI models. Compared to traditional AI methods, GAI has the advantage of being able to improve data by generating new data based on learned or inferred patterns. This process can expand the training set, which helps improve model generalization and address the problem of dataset sparseness. This is exactly what we will exploit in the following. First, the communication channel state is predicted, which is then used to generate channel-related parameters, such as loss versus transaction size.

*Predicting packet size given signal loss and power*

If the dependences of losses on packet sizes in Fig. 3 are taken not for a linear scale, but for a logarithmic one, then the graphs of losses y (%) on packet sizes x (bit) for different Power have the following form (Fig. 7):

y1 = 6.8 ln(x) +26 for Power = 5 dBm,
y2 = 7.1 ln(x) + 4 for Power = 7 dBm,
y3 = 6.2 ln(x) - 6 for Power = 9 dBm.

Using the calculated dependencies, we generate a loss data array with a step of 10 bits for x in the range from 10 to 10000 bits. We create a RandomForestRegressor model that is trained on this array to predict x given y (%) and Power. The code for the corresponding program is given in Appendix 2.

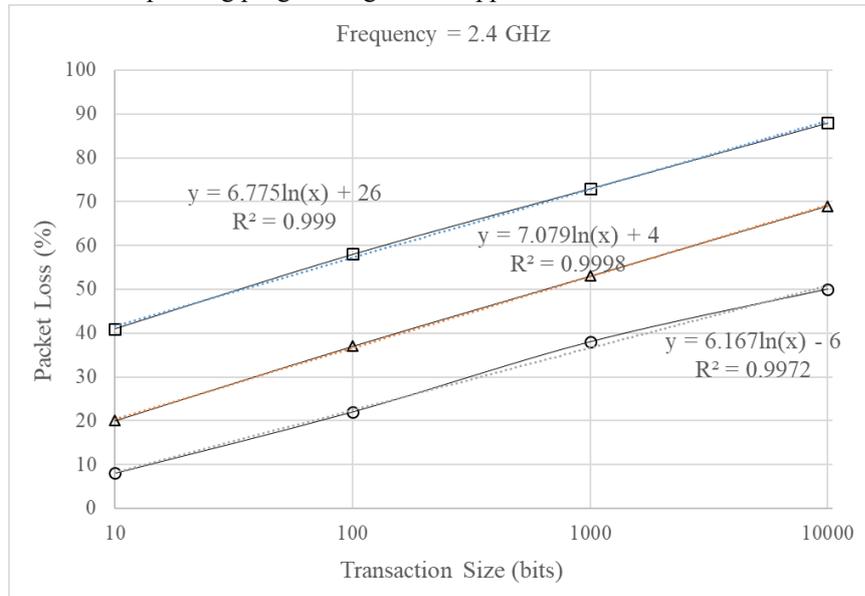

**Fig. 7.** Dependences of Packet Loss on TS for different transmit power for a logarithmic scale

*Adapting transmission parameters*

Let us assume that we want to reduce the predicted 46% loss level for a 20-bit packet at 5 dBm by increasing the power of the transmitted signal. We set the loss thresholds at 50% and 40%, upon reaching which the power should increase. Data transmission at Power = 5 dBm begins with 20 bits and increases in 10-bit increments every minute of transmission. The thresholds for the corresponding power levels are checked for achievement. Upon reaching the power thresholds, x is reduced by 20 bits, and transmission continues with the new power.

The procedure for adapting the channel parameters is shown in the graphs in Fig. 8-10, and the code for performing the described task is given in Appendix 3.

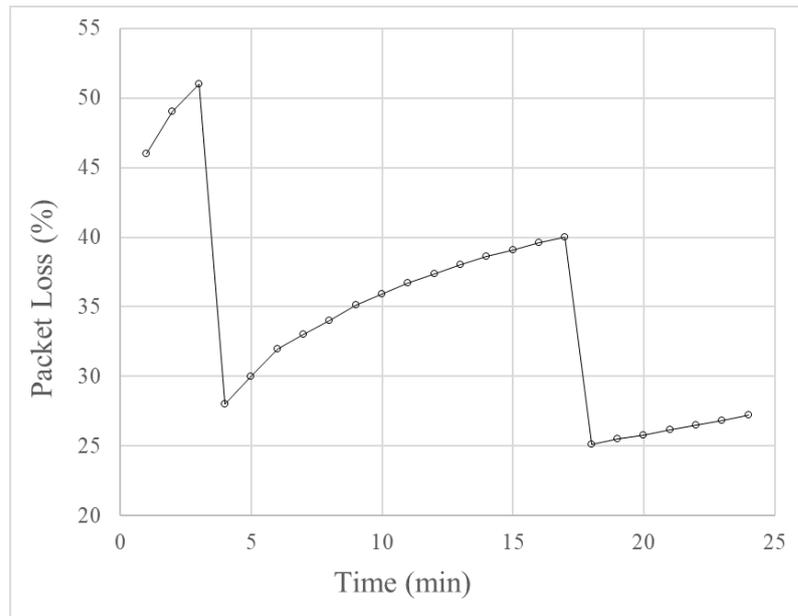

**Fig. 8.** Dependences of Packet Loss on time during AI adaptation

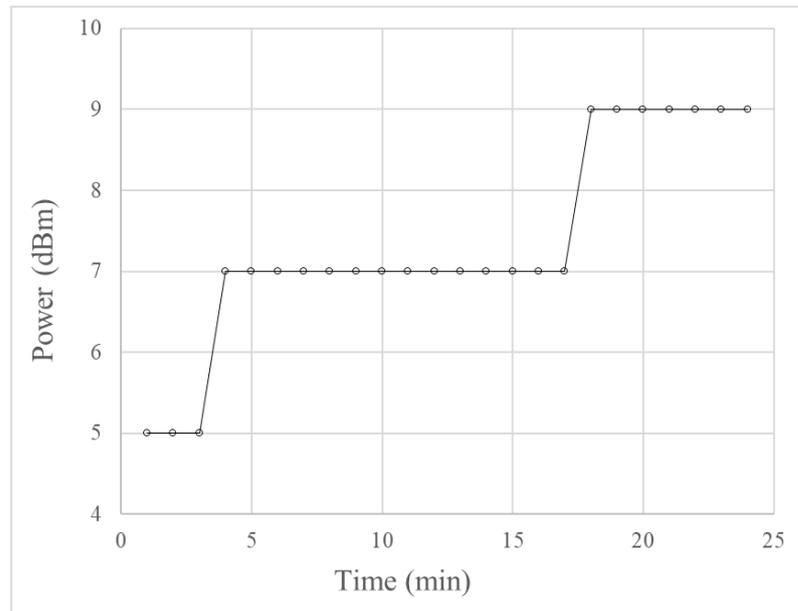

**Fig. 9.** Dependences of Power on time during AI adaptation

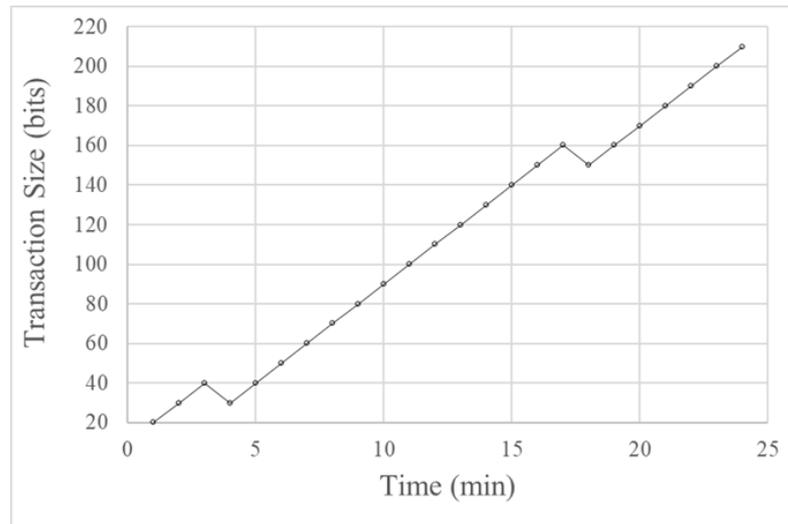

**Fig. 10.** Dependences of Transaction Size on time during AI adaptation

**Conclusion**
There are several takeaways from our study:
 1. Applicability of AI in UAV links.
 AI has proven to be a viable tool for adapting UAV links. It can dynamically adjust parameters such as packet size, power, and frequency to optimize performance. This is based on the demonstration of AI-driven adaptation mechanisms in the paper, showing real-time dependences on packet loss and other metrics.
 2. Insights into Packet Loss Behavior.
 Packet loss is highly dependent on factors such as transmit power, frequency, UAV flight area, and UAV count. Our study explicitly analyzes and provides dependences on these parameters, allowing for a deeper understanding and optimization of network performance.
 3. Importance of transmit power and frequency.
 Transmit power and operating frequency play a critical role in determining packet loss. Our results show how different power and frequency levels affect the reliability and efficiency of data transmission.
 4. Impact of mobility and network topology.
 The size of the UAV flight area and the number of UAVs significantly affect network performance. Large areas or more UAVs may lead to communication issues, increased routing overhead, or interference affecting packet loss and throughput.
 5. Practical implementation of adaptive mechanisms.
 The implementation of adaptive data transfer in software code demonstrates a practical way to improve UAV communications in the real world. By including the code, the study provides not only theoretical ideas but also a basis for further experimentation and application.
 6. Temporal dynamics during AI adaptation.
 AI adaptation effectively balances packet loss, power, and transaction size over time, indicating robust dynamic control. The time dependence of these parameters during AI adaptation highlights the ability of AI systems to dynamically respond to network conditions.
 The importance of our results is that they can help in designing more resilient and efficient UAV networks, especially in mission-critical applications. The role of AI in future networks is related to the demonstrated capabilities of AI in adapting network parameters, which strengthens its potential in future autonomous systems.
 Practical recommendations for UAV deployment are related to understanding packet loss under different conditions, which helps network engineers make informed decisions about deployment strategies. Our findings will help advance research and applications in UAV networks, especially in scenarios requiring adaptive and reliable communications.

**Appendix 1. Packet loss rate for different transaction sizes using the Friis loss model**
```
import numpy as np
import matplotlib.pyplot as plt
import random
import pandas as pd
#Constants
AREA_WIDTH = 1500
```

```python
AREA_HEIGHT = 1500
NUM_UAVS = 20
TX_POWER_DBM = 7
TX_POWER_MW = 10 ** (TX_POWER_DBM / 10)
NOISE_DBM = -100 # Typical noise floor in dBm
BANDWIDTH = 2e6 # 2 Mbps
FREQ = 2.4e9 # 2.4 GHz Wi-Fi
C = 3e8 # Speed of light
PACKET_SIZES = [10, 100, 1000, 10000] # in bits
NUM_PAIRS = 10
# Free-space path loss (Friis model)
def friis_path_loss(distance, freq):
    wavelength = C/freq
    return (wavelength / (4 * np.pi * distance)) ** 2
# Simulate UAV positions
np.random.seed(42) # For reproducibility
random.seed(42)
uav_positions = {
    i: (random.uniform(0, AREA_WIDTH), random.uniform(0, AREA_HEIGHT))
    for i in range(NUM_UAVS)
}
# Generate source-destination pairs
pairs = random.sample([(i, j) for i in range(NUM_UAVS) for j in range(NUM_UAVS) if i != j], NUM_PAIRS)
# Compute packet loss for each transaction size
def compute_packet_loss(packet_size, pairs):
    losses = []
    for src, dest in pairs:
        src_pos = np.array(uav_positions[src])
        dest_pos = np.array(uav_positions[dest])
        distance = np.linalg.norm(src_pos - dest_pos)
        if distance == 0:
            path_loss = 0
        else:
            path_loss = friis_path_loss(distance, FREQ)
        rx_power_mw = TX_POWER_MW * path_loss
        rx_power_dbm = 10 * np.log10(rx_power_mw) if rx_power_mw > 0 else -np.inf
        snr_db = rx_power_dbm - NOISE_DBM
        snr_linear = 10 ** (snr_db / 10)
        ber = 0.5 * np.exp(-snr_linear / 2) # Simplified BER for binary signaling
        loss_prob = 1 - (1 - ber)** packet_size
        losses.append(loss_prob * 100) # Convert to percentage
    return np.mean(losses)
# Data collection
results = []
for packet_size in PACKET_SIZES:
    avg_loss = compute_packet_loss(packet_size, pairs)
    results.append((packet_size, avg_loss))
# Convert results to a DataFrame
results_df = pd.DataFrame(results, columns=["Packet Size (bits)", "Packet Loss (%)"])
#Plotting
plt.figure(figsize=(10, 6))
plt.plot(results_df["Packet Size (bits)"], results_df["Packet Loss (%)"], marker="o", label="Packet Loss")
plt.xlabel("Packet Size (bits)")
plt.ylabel("Packet Loss (%)")
plt.title("Packet Loss vs. Packet Size")
plt.grid(True)
plt.legend()
plt.show()
# UAV initial positions plot
plt.figure(figsize=(10, 6))
for uav_id, (x, y) in uav_positions.items():
    plt.scatter(x, y, label=f"UAV {uav_id}")
```

```
plt.xlabel("X-coordinate (m)")
plt.ylabel("Y-coordinate (m)")
plt.title("Initial UAV Positions")
plt.grid(True)
plt.legend(bbox_to_anchor=(1.05, 1), loc='upper left')
plt.show()
# Print results table
print("Initial UAV Positions:")
for uav_id, pos in uav_positions.items():
 print(f"UAV {uav_id}: {pos}")
print("\nPacket Loss Data:")
print(results_df)
```

**Appendix 2. Predicting packet size given signal loss and power**
```
import numpy as np
import pandas as pd
from sklearn.preprocessing import PolynomialFeatures
from sklearn.ensemble import RandomForestRegressor
from sklearn.pipeline import make_pipeline
import matplotlib.pyplot as plt
# Defined dependencies
def y1(x): return 6.8 * np.log(x) + 26
def y2(x): return 7.1 * np.log(x) + 4
def y3(x): return 6.2 * np.log(x) - 6
# Generating data
x_values = np.arange(10, 10001, 10) # Batch sizes
data = []
for x in x_values:
data.append([x, y1(x), 5])
data.append([x, y2(x), 7])
data.append([x, y3(x), 9])
# Create DataFrame
df = pd.DataFrame(data, columns=["x", "y", "Power"])
df = df[df["y"] > 0] # Remove rows with negative loss values
# Training data
X = df[["y", "Power"]]
y = df["x"]
# Create model
model = make_pipeline(PolynomialFeatures(degree=2, include_bias=False), RandomForestRegressor(random_state=42))
model.fit(X, y)
# Test data
test_power = 9
test_y = 20
predicted_x = model.predict([[test_y, test_power]])
print(f"Predicted packet size x for y={test_y}% and Power={test_power} dBm: {predicted_x[0]:.2f} bits")
# Plot
plt.figure(figsize=(10, 6))
plt.plot(x_values, y1(x_values), label="Power = 5 dBm", color='blue')
plt.plot(x_values, y2(x_values), label="Power = 7 dBm", color='green')
plt.plot(x_values, y3(x_values), label="Power = 9 dBm", color='red')
plt.scatter([predicted_x], [test_y], color='black', label="Prediction", zorder=5)
plt.xlabel("Packet size (bit)")
plt.ylabel("Losses (%)")
plt.title("Losses vs. packet size for different powers")
plt.legend()
plt.grid()
plt.show()
```

**Appendix 3. Code for adapting transmission parameters**
```
import numpy as np
import matplotlib.pyplot as plt
```

```python
# Function for calculating losses y depending on x and power
def calculate_loss(x, power):
    if power == 5:
        return 6.8 * np.log(x) + 26
    elif power == 7:
        return 7.1 * np.log(x) + 4
    elif power == 9:
        return 6.2 * np.log(x) - 6
    else:
        raise ValueError("Invalid power value")
# Initializing variables
x = 20 # Initial packet size (bit)
time = 0 # Time (minutes)
power = 5 # Initial power value (dBm)
x_history = [] # Packet size history
y_history = [] # Loss history
power_history = [] # Power history
time_history = [] # Time history
# Stop condition
threshold_reached = False
# Main parameter adaptation loop
while not threshold_reached:
    y = calculate_loss(x, power) # Loss calculation
    # Write current values to history
    x_history.append(x)
    y_history.append(y)
    power_history.append(power)
    time_history.append(time)
    # Output current state
    print(f"Time: {time} min, Packet size: {x} bits, Loss: {y:.2f}%, Power: {power} dBm")
    # Check for thresholds reached
    if power == 5 and y >= 50:
        print(f"Adaptation threshold reached at Power = 5 dBm, x = {x} bits")
        x -= 20 # Decrease packet size
        power = 7 # Increase power
    elif power == 7 and y >= 40:
        print(f"Adaptation threshold reached at Power = 7 dBm, x = {x} bits")
        x -= 20 # Decrease packet size
        power = 9 # Increase power
    elif power == 9 and y >= 30:
        print(f"Adaptation threshold reached at Power = 9 dBm, x = {x} bits")
        threshold_reached = True # Termination condition
    # Increase packet size and time
    x += 10
    time += 1
# Plotting graphs
plt.figure(figsize=(12, 8))
# Plot y(t)
plt.subplot(3, 1, 1)
plt.plot(time_history, y_history, marker="o", label="y(t)")
plt.axhline(50, color="red", linestyle="--", label="Threshold 50%")
plt.axhline(40, color="green", linestyle="--", label="Threshold 40%")
plt.axhline(30, color="blue", linestyle="--", label="Threshold 30%")
plt.xlabel("Time (min)")
plt.ylabel("Loss y (%)")
plt.legend()
plt.grid()
# Power(t) graph
plt.subplot(3, 1, 2)
plt.plot(time_history, power_history, marker="o", color="orange", label="Power(t)")
plt.xlabel("Time (min)")
```

```
plt.ylabel("Power (dBm)")
plt.legend()
plt.grid()
# Plot x(t)
plt.subplot(3, 1, 3)
plt.plot(time_history, x_history, marker="o", color="purple", label="x(t)")
plt.xlabel("Time (min)")
plt.ylabel("Packet size x (bits)")
plt.legend()
plt.grid()
plt.tight_layout()
plt.show()
# Output data frame
import pandas as pd
data = pd.DataFrame({
"Time (min)": time_history,
"Packet size x (bits)": x_history,
"Loss y (%)": y_history,
"Power (dBm)": power_history
})
print("\nData table:")
print(data)
```

**Statements & Declarations**
**Funding**
The authors declare that no funds, grants, or other support were received during the preparation of this manuscript.
**Competing Interests**
The authors have no relevant financial or non-financial interests to disclose.
**Conflicts of Interest**
The authors declare no conflict of interest.
**Author Contributions**
Volodymyr Kharchenko – V.Kh., Andrii Grekhov – A.G., Vasyl Kondratiuk – V.K.
*Conceptualization*, A.G. and V.Kh.; *methodology*, A.G.; *validation*, A.G., V.Kh. and V.K.; *investigation*, A.G.; *resources*, V.Kh. and V.K.; *writing*—original draft preparation, A.G.; *writing*—review and editing,V.K.; *supervision*, V.Kh.; *project administration, V.K.*; All authors have read and agreed to the published version of the manuscript.
 **Ethics approval**
Not applicable.
**Data Availability Statement**
All data generated and analyzed during this study are included in this article. The datasets generated during the current study are available from the corresponding author on request.